\documentclass[12pt]{article}
\usepackage{psfrag}
\usepackage{graphicx}
\usepackage{epstopdf}
\usepackage{amsmath}
\usepackage[T1]{fontenc}
\usepackage[utf8]{inputenc}

\begin{document}

\title{Importance of thermodynamic fluctuations in the Gaździcki Gorenstein model}
\date{\today}
\author{Kacper Zalewski\\Institute of Nuclear Physics, Krakow, \\ Institute of Physics Jagellonian University}
\maketitle
\begin{abstract}
Effects of the standard thermodynamic fluctuations on the predictions of the Gaździcki Gorenstein model of particle production in high-energy heavy ions collisions are evaluated. At low numbers of participating nucleons the corrections due to these fluctuations are found to be very significant.
\end{abstract}

\section{Introduction}

Gaździcki and Gorenstein  \cite{GAZ} proposed a description of the fluid produced in high-energy heavy ion collisions, where the fluid contains at low energies only hadrons ($W$-phase) and at high energies only partons ($Q$-phase). In the following, both hadrons and partons will be referred to as particles.   The two phases can coexist. For plausible values of the parameters of the model,  there is a first order phase transition, similar to the familiar water-vapour transition, at collision energies per pair of colliding nucleons $\sqrt{s_{NN}} \approx 9$GeV. At the low energy end of the transition region the model predicts a maximum in the ratio of the number of produced $K^+$ mesons to the number of produced $\pi^+$ mesons. This maximum has indeed been found. For a compilation of the experimental data see \cite{RUS}.

In the original approach of Gaździcki and Gorenstein  the thermodynamic limit was used \cite{ZAL}. Then, approximate strangeness conservation follows from assuming the same chemical potentials for the corresponding strange and anti-strange particles. In the model \cite{GAZ} the chemical potentials of all the particles are zero.  In a subsequent paper \cite{PGG}, Poberheznyuk, Gaździcki and Gorenstein imposed exact strangeness conservation. This goes beyond the thermodynamic limit and yields some unfamiliar results. E.g., when going from the W-phase to the Q-phase the temperature slightly decreases with increasing energy.

In the thermodynamic limit, at given values of the total volume, energy and all the chemical potentials, the fraction $\lambda V$ of the total volume $V$ occupied by phase $Q$ is unambiguously defined and corresponds to the maximum of entropy. Since by definition

\begin{equation}\label{limlan}
  0 \leq \lambda \leq 1,
\end{equation}
this maximum can be at $\lambda=0$ ($W$-phase), $\lambda = 1$ ($Q$ phase) or somewhere in between (coexistence of the two phases).

In the standard theory of thermodynamic fluctuations (Einstein 1907) one assumes that the parameter $\lambda$ has the probability distribution

\begin{equation}\label{pralam}
  \rho(\lambda) = C e^{S(\lambda)},
\end{equation}
where the dependence on the parameters other than $\lambda$ has not been written explicitly. The thermodynamic limit corresponds to the replacement of this distribution by a Dirac $\delta$-distribution. In the present paper we consider the effects, on the predictions of the model, of using formula (\ref{pralam}) with the known entropy $S(\lambda)$, instead of the $\delta$-distribution. Then all the values of $0 \leq \lambda \leq 1$ are possible at any energy though, at each energy, some of them have very small probabilities. In other words, we will discuss the effects of the thermodynamical fluctuations of the volume fraction $\lambda$.

A slightly simplified version of the model from \cite{GAZ} will be used. This makes it possible to write many of the results in simple analytic forms, while it does not change the qualitative conclusions.

\section{The model}

In the model from \cite{GAZ} the overall collision energy fixes the energy of the fluid

\begin{equation}\label{}
  E = A_p\eta(\sqrt{s_{NN}} - 2 m)
\end{equation}
and its volume

\begin{equation}\label{forvol}
  V = \frac{A_p}{\rho_0}\frac{2m}{\sqrt{s_{NN}}}.
\end{equation}
In these formulae $m$ is the nucleon mass,

\begin{equation}\label{}
  \rho_0 = 0.16\mbox{fm}^{-3}
\end{equation}
is the rest frame nuclear density,

\begin{equation}\label{}
  \eta = 0.67
\end{equation}
is a phenomenological factor correcting for the energy, in excess of $2m$, taken away by the leading particles, which should not be included into the energy of the fluid. $A_p$ is the number of interacting nucleons from one nucleus. For simplicity it has been assumed that $A_p$ is the same for each of the two colliding nuclei.

The fluid is an ideal gas, except that in the $Q$-phase a term $\lambda VB$, where $B$ is the bag constant, is added to the energy. The (anti)strange particles in the $Q$ phase have mass $175$MeV and in the $W$ phase $500$MeV. All the remaining particles are massless. In the somewhat simplified version of the model described in \cite{PGG} Boltzmann statistics is used for all the (anti)strange particles and Bose-Einstein statistics for all the non-strange ones.

Since our purpose is to demonstrate the importance of the thermodynamic fluctuations of the volume fraction $\lambda$, and not a quantitative comparison with the data, we choose the simplest version of the model, with all the particles massless and subject to Boltzmann statistics. The assumption that all the particles are massless has been used in \cite{GAZ} for illustrative purposes. The replacement of quantum statistics by the Boltzmann one (for the (anti)strange particles) is one of the differences between the approaches in \cite{PGG} and in \cite{GAZ}.

For the effective numbers of states for non-strange (ns) and (anti)strange (s) particles at given momentum we choose

\begin{equation}\label{cupcon}
 g_{Wns}=17.31;\qquad g_{Ws}=8.01;\qquad g_{Qns} = 43.29;\qquad g_{Qs} = 10.78.
\end{equation}
These numbers were obtained from the corresponding numbers given in \cite{PGG} by multiplying the numbers for the non-strange particles by $\frac{\pi^4}{90}$, in order to correct for the change of statistics, and by multiplying the numbers for the (anti)strange particles by factors which compensate, at temperature $T = 200$MeV, the effects of the changes of mass in the contributions to the energies of the two phases. Analogous corrections were used in \cite{PGG} to compare their numbers of states with those from \cite{GAZ}.

These assumptions imply that the grand canonical potential is

\begin{equation}\label{omega}
  \Omega = -g(\lambda )Tze^{-\beta\mu} + \lambda BV.
\end{equation}
In this formula $\beta$ is the inverse temperature $\frac{1}{T}$, $\mu$ is the chemical potential,  assumed to be the same for all the particles,

\begin{equation}\label{}
  g(\lambda) = g_W + \lambda(g_Q-g_W)
\end{equation}
with $g_W = g_{Wns} + g_{Ws}$ and $g_Q = g_{Qns} + g_{Qs}$ and the single particle phase space

\begin{equation}\label{defzed}
  z = \frac{VT^3}{\pi^2}.
\end{equation}

Multiplying the potential $\Omega$ by $\beta$ and differentiating the result with respect to $\beta$, at constant $V$ and $\mu \equiv 0$,  we obtain the energy of the fluid

\begin{equation}\label{energy}
  E = 3Tg(\lambda)z + \lambda BV.
\end{equation}
Introducing the dimensionless energy density

\begin{equation}\label{epsbar}
  \overline{\epsilon} = \frac{E}{BV},
\end{equation}
using (\ref{defzed}) and (\ref{energy}) we obtain

\begin{equation}\label{formaz}
  z = \frac{VB^{\frac{3}{4}}}{\sqrt{\pi}}\left(\frac{\overline{\epsilon}-\lambda}{3g(\lambda)}\right)^{\frac{3}{4}}.
\end{equation}

The pressure is

\begin{equation}\label{forpre}
  p = -\frac{\Omega}{V} = \frac{B}{3}(\overline{\epsilon} - 4\lambda)
\end{equation}
and the temperature

\begin{equation}\label{fortem}
  T = \left(\frac{\pi^2 B(\overline{\epsilon}-\lambda)}{3g(\lambda)}\right)^\frac{1}{4}.
\end{equation}
Since the chemical potentials vanish, the entropy

\begin{equation}\label{forent}
  S = \frac{E-\Omega}{T} = 4g(\lambda)z.
\end{equation}

In order to calculate the average numbers of non-strange and (anti)strange particles it is necessary to split the potential $\Omega$ into the two corresponding contributions. This is done by making the replacement

\begin{equation}\label{}
  g(\lambda) = g_{ns}(\lambda) + g_s(\lambda),
\end{equation}
where

\begin{equation}\label{}
  g_i(\lambda) = g_{Wi} + (g_{Qi} - g_{Wi})\lambda, \qquad i = ns,s.
\end{equation}
Differentiating the two terms in the potential $\Omega$ with respect to $\mu$, putting $\mu=0$ and changing signs one finds

\begin{equation}\label{}
  N_i = g_i(\lambda)z,\qquad i = ns,s
\end{equation}
and for the ratio of the average numbers of (anti)strange and non-strange particles

\begin{equation}\label{ratnum}
  \frac{N_s}{N_{ns}} = \frac{g_s(\lambda)}{g_{ns}(\lambda)}.
\end{equation}
Outside the coexistence region these ratios are constant, due to the assumption that all the particles have equal masses. This could be easily corrected by giving suitable masses to the (anti)strange particles \cite{GAZ}. We chose the simpler version of the model, because it is more transparent, while for our discussion it is good enough.

\section{Thermodynamic limit}

In the thermodynamic limit $\lambda$ maximizes the entropy (\ref{forent}) under condition (\ref{limlan}). Equating to zero the derivative of the entropy with respect to $\lambda$ we get

\begin{equation}\label{lamter}
  \lambda = \frac{1}{4}(\overline{\epsilon} - 3\overline{g}),
\end{equation}
where

\begin{equation}\label{}
  \overline{g} = \frac{g_W}{g_Q-g_W}.
\end{equation}
According to condition (\ref{limlan}), this relation can be used only for

\begin{equation}\label{coepha}
 3\overline{g} \leq \overline{\epsilon} \leq 3\overline{g} +4.
\end{equation}
When  $\overline{\epsilon}< 3\overline{g}$, the maximum entropy corresponds to $\lambda =0$, i.e. the system is in the $W$-phase. When $\overline{\epsilon}> 3\overline{g} + 4$, the maximum entropy corresponds to $\lambda = 1$ and the system is in the $Q$-phase. In order to find the corresponding limits for the energy $\sqrt{s_{NN}}$ it is necessary to know the bag constant $B$.

In the range (\ref{coepha}) relation (\ref{fortem}) yields

\begin{equation}\label{temter}
  T = \left(\frac{\pi^2B}{g_Q-g_W}\right)^\frac{1}{4}.
\end{equation}
Thus, in the coexistence region the temperature is constant. Assuming \cite{GAZ} that there it equals $200$MeV, one finds

\begin{equation}\label{}
  B = 607\mbox{MeV fm}^{-3}.
\end{equation}
The energy range of the coexistence region is, therefore,

\begin{equation}\label{eneran}
  6.33\mbox{GeV} \leq \sqrt{s_{NN}} \leq 9.40\mbox{GeV}.
\end{equation}

Substituting the solution (\ref{lamter}) for $\lambda$ into formula (\ref{forpre}) we get for the pressure in the transition region

\begin{equation}\label{}
  p = \overline{g}B = 534\mbox{MeVfm}^{-3}.
\end{equation}
Thus also the pressure is constant there.

As seen from (\ref{ratnum}), in the transition region the ratio of the average number of (anti)strange particles to the average number of the non-strange particles drops from

\begin{equation}\label{}
  \frac{g_{Ws}}{g_{Wns}} = 0.46
\end{equation}
at $\lambda = 0$ i.e. $\sqrt{s_{NN}} = 6.33$GeV, to

\begin{equation}\label{}
  \frac{g_{Qs}}{g_{Qns}} = 0.25
\end{equation}
at $\lambda = 1$ i.e. $\sqrt{s_{NN}} = 9.40$GeV. This is the high-energy side of the "horn" observed in experiment.

\section{Thermodynamic fluctuations}

Let us consider now the effects of the thermodynamic fluctuations in the volume fraction $\lambda$ on the average values of the parameters of the system. The corrected averages are obtained by averaging the $\lambda$-dependent values obtained in Section 2. over the distribution (\ref{pralam}) with the entropy given by (\ref{forent}). According to the general rules of statistical thermodynamics, at high $A_p$ the thermodynamic results should be reproduced. Thus the interesting questions are: what happens at low values of $A_p$ and how fast the thermodynamic limit is reached.

It is instructive to begin with the parameter $\lambda$. The results are shown in Fig. 1. The broken line (green on line) corresponds to the thermodynamic limit, the line close to it (blue on line) has been calculated putting $A_p =10$ and the remaining line (red on line)  corresponds to $A_p=1$, which is the case discussed in \cite{PGG}.

\begin{figure}[tpb!]
\centering
\includegraphics{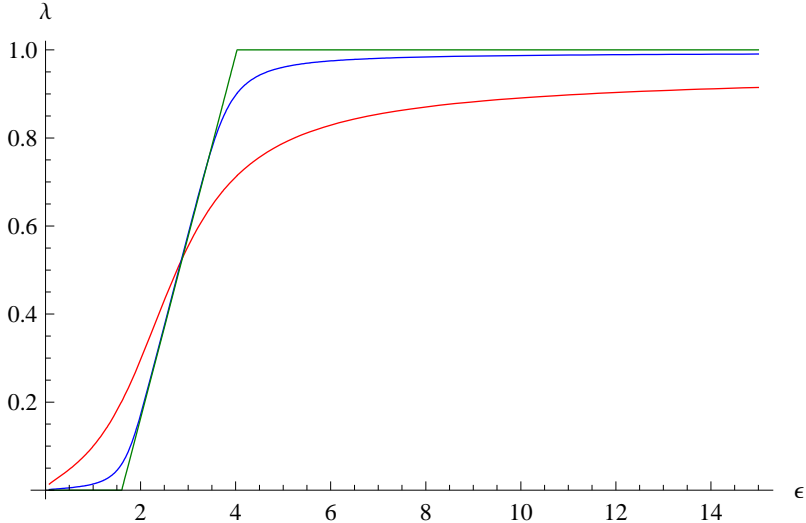}
\caption{Dependence of the average volume fraction $\lambda$ on the energy density $\epsilon = \frac{E}{V}$. For the meaning of the lines see text.}
\end{figure}
It is seen that at $A_p = 1$, the region where the two phases can coexist is greatly extended, as
compared to the thermodynamic limit. Moreover, in the thermodynamic limit of the pressure the energy dependence of the ideal gas term is exactly cancelled by the energy dependence of the term proportional to the bag constant; the thermodynamic fluctuations in the parameter $\lambda$ affect the second term, but not the first one, therefore, the cancellation is no more expected. The numerical results for the pressure are shown in Fig.~2. At $A_p=1$ the plateau in $p$ is hardly visible, while at $A_p = 10$ the thermodynamic limit is a very good approximation.

\begin{figure}[tpb]
\centering
\includegraphics{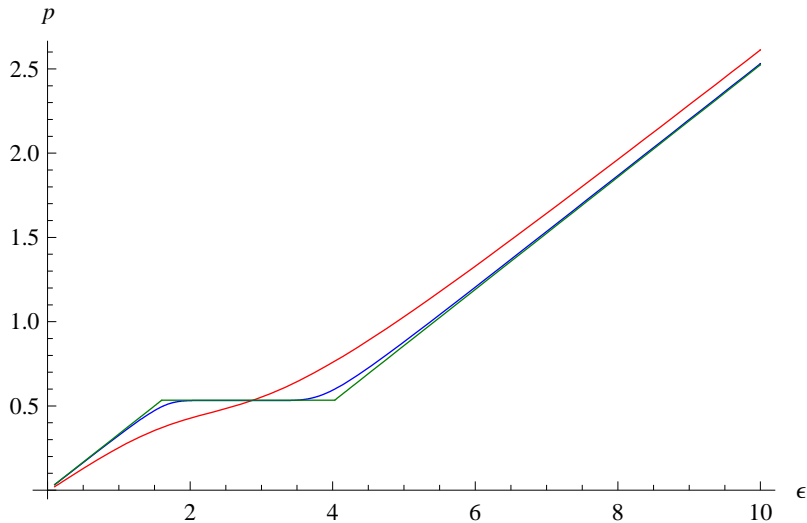}
\caption{Dependence of the average pressure $p$ on the energy density $\epsilon = \frac{E}{V}$.  The meaning of the lines as in Fig. 1.}
\end{figure}
The effect of thermodynamic fluctuations on the energy density dependence of the temperature is  qualitatively similar to that for the pressure. This is shown in Fig. 3.

\begin{figure}[tpb!]
\centering
\includegraphics{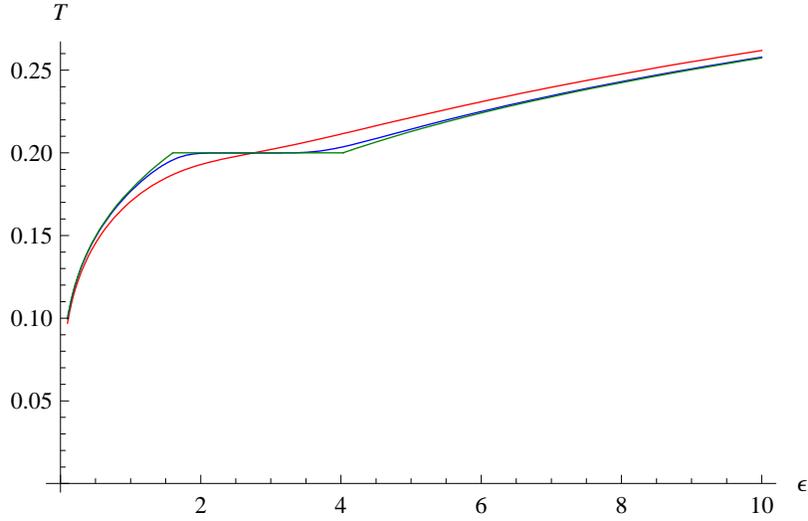}
\caption{Dependence of the average temperature $T$ on the energy density $\epsilon = \frac{E}{V}$. The meaning of the lines as in Fig. 1.}
\end{figure}

Finally, the ratio of the number of strange particles to the number of non strange particles is shown in Fig. 4. It is seen that the thermodynamic fluctuations at $A_p=1$ make the decrease of this ratio with increasing energy density significantly slower. Again at $A_p=10$ we are very close to the thermodynamic limit.

\begin{figure}[tpb!]
\centering
\includegraphics{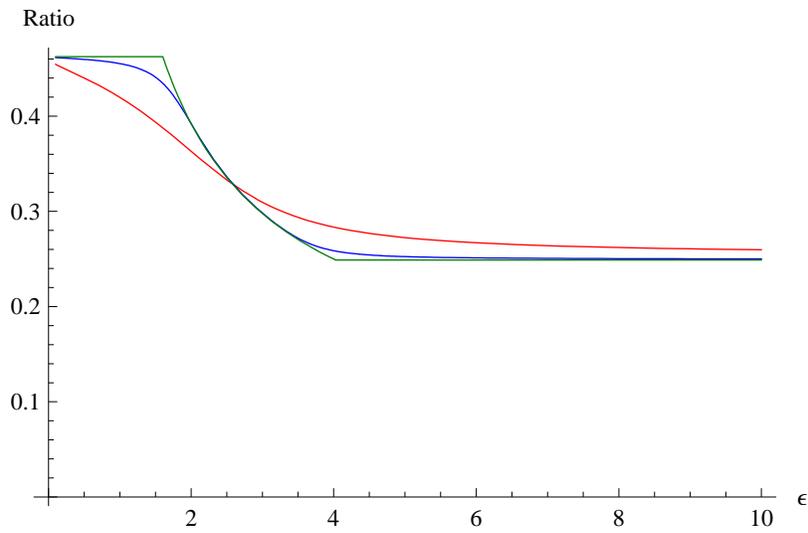}
\caption{Dependence of the average ratio of the number of strange particles to the number of nonstrange particles on the energy density $\epsilon = \frac{E}{V}$. The meaning of the lines as in Fig. 1.}
\end{figure}

\section{Exact strangeness conservation}

The version of the model \cite{GAZ} with strict strangeness conservation \cite{PGG} can be described as follows. The fluid of nonstrange particles is described as before. The fluid of (anti)strange particles consists of pairs of strangeness zero. The pairs are so loosely bound that the single pair phase space is just the square of the single particle phase space. The combinatorial factor for $N$ identical pairs is $\frac{1}{(N!)^2}$, which corrects for the unobservable permutations of the strange and of the antistrange particles. The corresponding grand partition function can be calculated in closed form \cite{PGG} and the potential $\Omega$ is

\begin{equation}\label{}
  \Omega = -Tg_s(\lambda)z - T\log I_0[g_s(\lambda)z] + \lambda BV,
\end{equation}
where $I_0$ is the modified Bessel function. We will use an approximate version of this formula, where the terms proportional to inverse powers of $A_P$ are neglected. Then

\begin{equation}\label{}
  \Omega = -Tg(\lambda)z + \lambda B V + \frac{1}{2}T\log(2\pi g_s(\lambda)z).
\end{equation}
This yields for the entropy

\begin{equation}\label{entssb}
  S = 4g(\lambda)z - \frac{1}{2}\log(2\pi g_s(\lambda)z) + \frac{3}{2}.
\end{equation}

Since the correction to the entropy is a decreasing function of the parameter $\lambda$, the maximum of the entropy is shifted towards lower values of this parameter. Because of this shift, at the energy densities where in the thermodynamic approximation $\lambda$ was zero or one, now it is, respectively, a little less than zero an a little less than one. In the transition region $\lambda$ is an increasing function of the energy density. Therefore, to go back to $\lambda = 0$ and $\lambda = 1$ it is necessary to increase the corresponding energy densities -- the transition region gets shifted towards higher energies. As seen from the formulae given in Section 2., the decrease of $\lambda$ at fixed energy density implies that in the transition region  the temperature and the pressure increase. The same results from the increase of the energy density at given $\lambda$. All these effects, however, are small because the $\lambda$-dependent parts of the corrections are by factors of order $A_p$ smaller than the main terms obtained in the thermodynamic limit.

In our approximation, formula (\ref{lamter}) gets replaced by

\begin{equation}\label{implam}
  \lambda = \frac{1}{4}(\overline{\epsilon} - 3\overline{g})- \frac{3\sqrt{\pi}}{2A_p}
  \frac{(\overline{g}_s-\overline{g})\rho_0}{(g_Q-g_W)^\frac{1}{4}B^\frac{3}{4}} \frac{\sqrt{s_{NN}}}{2m}\frac{1}{\overline{\epsilon} - 3\overline{g} + 4\overline{g}_s},
\end{equation}
where

\begin{equation}\label{}
  \overline{g}_s = \frac{g_{Ws}}{g_{Qs} - g_{Ws}}.
\end{equation}
In formula (\ref{implam}) the terms of higher order in $A_p^{-1}$ have been neglected. Comparison with the exact solution shows that down to $A_p=1$ this is a very good approximation.

Substituting (\ref{implam}) into the formulae from Section 2. we find the results, for $A_p=1$,  corresponding to those from \cite{PGG}. Since our calculation is analytic, the following conclusions, concerning the transition region, are easily checked.

\begin{itemize}
  \item Across the transition region the correction to the parameter $\lambda$ decreases from $-0.046A_p^{-1}$ to $-0.051A_p^{-1}$. As seen from Fig 1., for $A_p=1$, in most of the transition region, this is much less than the effect of replacing $\lambda$ by its average, as  discussed in Section 4. When $A_p$ increases the correction term in (\ref{implam}) is proportional to  $A_p^{-1}$, while the range of the thermodynamic fluctuations goes like $A_p^{-\frac{1}{2}}$, therefore, the relative importance of the correction term in  (\ref{implam}) decreases.  Nevertheless, this correction has some interesting implications.
  \item Since the correction to $\lambda$ is negative, the energies $\sqrt{s_{NN}}$ corresponding to the limits  of the transition region increase. This is a small effect: $0.18$GeV at the low energy end and $0.13$GeV at the high energy end.
  \item For the pressure the exact cancellation of the energy dependent terms does not hold any more, but it is still a good approximation. In our model the pressure is increased by $37$MeVfm$^{-3}$ at the beginning and by $41$MeVfm$^{-3}$ at the end of the transition region.
  \item As seen from (\ref{fortem}), at given $\lambda$, thus in particular at the ends of the transition region, the temperature increases with increasing $\overline{\epsilon}$ and, consequently, with increasing $\sqrt{s_{NN}}$. In our model the increase is $4$MeV at the beginning of the transition region  and $2$MeV  at the end.
\end{itemize}

These results are in qualitative agreement with the results obtained numerically in \cite{PGG} except for one point. The small shift in energy of the interaction region is positive according to our analysis, while it is negative in \cite{PGG}. In order to include the thermodynamic fluctuations of the volume fraction $\lambda$, one has to repeat the calculations from Section 4. using for the entropy expression (\ref{entssb}) instead of the expression (\ref{forent}). Since the results are very similar to those from Section 4., we do not give them here.

\section{Discussion and conclusions}

The thermodynamic limit is calculated by making the volume tend to infinity with all the intensive, i.e. measurable locally like the pressure or the temperature, parameters kept fixed. As seen from (\ref{forvol}), in the Gaździcki Gorenstein model the volume depends on the number of interacting nucleons $A_p$ and on the nucleon-nucleon collision energy $\sqrt{s_{NN}}$. Any change of
$\sqrt{s_{NN}}$ changes the energy density of the fluid, which is an intensive parameter. Therefore, the thermodynamic limit corresponds to $A_p$ tending to infinity at constant $\sqrt{s_{NN}}$. Of course, in experiment $A_p$ cannot exceed the number of nucleons in the colliding nucleus, but formally the limit can be taken and used to get predictions at finite $A_p$. This problem, as well as its solution, is well known, e.g. from the thermodynamics of ideal gases.

In the thermodynamic approximation the fraction $\lambda$ of the volume which is occupied by the $Q$-phase is a well-defined function of the collision energy. Our observation is that the fluctuations of $\lambda$ become important for $A_p$ close to one, though they are of little importance already for $A_p=10$. For the case considered in \cite{PGG}, i.e. for $A_p = 1$, including the fluctuations changes the picture significantly. The energy region where $\lambda$ equals neither zero nor one, i.e. the transition region, becomes much wider. The plateaus in the dependence of the temperature and pressure on the energy, which are characteristic for the phase transitions of the Van der Waals type, disappear. The decrease, with increasing energy, of the ratio of the number of strange particles to the number of non-strange particles in the transition region becomes much slower.

The entropy used in the theory of thermodynamic fluctuations is in the thermodynamic limit. Therefore, it is a function of state and the fact that the system is isolated is of no importance for it. Calculating the averages, however, it is important to include all the allowed states and no others. In the present paper the averaging is made at constant energy density and volume, i.e. at constant collision energy $\sqrt{s_{NN}}$. Moreover it is assumed that all the chemical potentials are equal zero.

When exact strangeness conservation is included the thermodynamic function of the fluid still can be calculated exactly \cite{PGG}. We prefer, however, to use an approximation obtained by omitting the terms of higher order in $A_p^{-1}$. At first sight it may seem surprising that this is a good approximation at $A_p=1$, but a comparison with the exact results shows that this is indeed the case. We have done the calculations both for the approximate and for the exact formulae. We chose for presentation the approximate results, because they give much more physical insight. Moreover,  the corrections due to exact strangeness conservation are small, so that not much is gained by calculating them more precisely.

\textbf{\emph{Acknowledgement}}
The author thanks Marek Gaździcki and Mark Gorenstein for helpful comments. This work was partly supported by the Polish National Science Center (NCN) under grant DEC-2013/09/B/ST2/00497.

\end{document}